\documentclass[preprint,notoc]{JHEP3}
\usepackage{epsfig}

\def\etmiss{E_T^{\rm miss}}
\def\eslt{E_T^{\rm miss}}

\def\to{\rightarrow}

\def\bi{\begin{itemize}}
 \def\ei{\end{itemize}}

\def\c1p{C1^\prime}

\def\tf{\tilde f}

\def\tg{\tilde g}

\def\tq{\tilde q}

\def\tz{\widetilde Z}
\def\alt{\stackrel{<}{\sim}}

\def\be{\begin{equation}}  
\def\ee{\end{equation}}  
\def\bea{\begin{eqnarray}}  
\def\eea{\end{eqnarray}}  

\def\sps1ap{SPS1a$^\prime$}
\title{Capability of LHC to discover supersymmetry\\
with $\sqrt{s}=$7 TeV and 1 fb$^{-1}$
}
\author{Howard Baer$^{a}$, Vernon Barger$^b$, Andre Lessa$^a$ 
and Xerxes Tata$^{b,c}$\\
$^a$Dept.\ of Physics and Astronomy, University of Oklahoma, Norman, OK 73019, USA\\
$^b$Dep't of Physics, University of Wisconsin, Madison, WI 53706, USA\\
$^c$Dept. of Physics and Astronomy, University of Hawaii, Honolulu, HI 96822, US\\
E-mail: \email{baer@nhn.ou.edu}, \email{barger@pheno.wisc.edu},
\email{lessa@nhn.ou.edu}, \email{tata@phys.hawaii.edu}}

\preprint{\vbox{}}

\abstract{ We examine the capability of the CERN Large Hadron Collider to 
discovery supersymmetry (SUSY) with energy $\sqrt{s}=7$ TeV and integrated 
luminosity of about 1 fb$^{-1}$.
Our results are presented within the paradigm minimal supergravity model
(mSUGRA or CMSSM).
Using a 6-dimensional grid of cuts for optimization of
signal to background-- including missing $E_T$-- 
we find for $m_{\tg}\sim m_{\tq}$ an LHC reach of 
$m_{\tg}\sim 800,\ 950,\ 1100$ and 1200 GeV
for 0.1, 0.3, 1 and 2 fb$^{-1}$, respectively.
For $m_{\tg}\ll m_{\tq}$, the reach is instead near 
$m_{\tg}\sim 480,\ 540,\ 620$ and 700 GeV, for the same
integrated luminosities.
We also examine the LHC reach in the case of very low integrated 
luminosity where missing $E_T$  may not be viable. 
We focus on the multi-muon, multi-lepton (including electrons) and
dijet signals.
Although the LHC reach without $\eslt$ is considerably lower
in these cases, it is still substantial: 
for 0.3 fb$^{-1}$, the dijet reach in terms of gluino mass 
is up to 600 GeV for very low $m_0$, while the dilepton reach is to 
gluino masses of $\sim 500$ GeV over a range of $m_0$ values.
}  
\begin{document}

\section{Introduction}
\label{sec:intro}

The CERN Large Hadron Collider (LHC) has recently begun to generate data from
proton-proton collisions at $\sqrt{s}= 7$~TeV. 
The plan is to run for much of the next two years, 
with a goal of accumulating $\sim 1$ fb$^{-1}$ of usable data. 
This initial run will be followed by a shut down for a year or so for various
upgrades, followed by a turn-on at or near design energy of 
$\sqrt{s}=14$ TeV.

The discovery capability of LHC with $\sqrt{s}=14$ TeV (LHC14), has been
investigated for several new physics scenarios, where supersymmetry
(SUSY) \cite{wss} is frequently used as a canonical example
\cite{lhcreach}.  In the paradigm minimal supergravity (mSUGRA or CMSSM)
model \cite{msugra} based on local supersymmetry \cite{local}, the LHC14
reach with 100 fb$^{-1}$ was found to extend to $m_{\tg}\sim 3.1$ TeV
for $m_{\tq}\sim m_{\tg}$, and to $m_{\tg}\sim 1.8$ TeV, for $m_{\tq}\gg
m_{\tg}$.

As LHC turn-on drew near, the question turned to how well LHC could do
in its initial stages, at very low integrated luminosity, and perhaps
before the LHC detectors are fully calibrated, so that the canonical SUSY
signature -- the presence of mult-jet plus large missing $E_T$
($\eslt$) --
is not fully viable. In Ref.~\cite{early,rts,lhc10}, it was emphasized that
SUSY could be discovered at LHC even without using $\eslt$, by focusing
instead on events with large multiplicity of isolated leptons.
In Ref.~\cite{rts}, it was shown that LHC could discover SUSY in the dijet
channel, using new kinematic variables, even without viable $\eslt$.

In a previous study \cite{lhc10}, we investigated the supersymmetry
discovery potential of LHC with $\sqrt{s}=10$ TeV (LHC10), the energy at
which the machine was then expected to operate, both with and without
the use of $\eslt$, and compared it to the reach of LHC14.  After
Ref.~\cite{lhc10} appeared, the decision was made to operate LHC at half
its design energy of $\sqrt{s}=7$ TeV (LHC7).  Furthermore, the
additional year of LHC down-time allowed the various detectors to amass
millions of cosmic muon events. This array of cosmic data allowed the
experiments to make progress on important issues of detector alignment,
tracking and calibration.  At the end of 2009, the first proton-proton
collisions were recorded in the CMS and ATLAS (and ALICE and LHC-b)
detectors at center-of-mass energies of 900 GeV and 2.36 TeV. Initial
analyses of these events show remarkably good agreement between Monte
Carlo expectations and the actual data, including the (very low energy)
$\eslt$ spectrum \cite{cms,atlas}.  By March 30, 2010, the first $pp$
collisions were recorded at $\sqrt{s}=7$ TeV.  At this time, millions of
$pp$ collision events at 7 TeV have been recorded, including various
multi-jet events, and even candidate leptonically decaying $W$
events.
%
%(after noise cleaning algorithms that reduced the large $\eslt$ tail by about a factor 
%of 5) for which the
%CMS collaboration uses the track-corrected and particle flow $\eslt$

In light of the CERN decision to perform a major collider run at
$\sqrt{s}=7$ TeV with $\sim 1$ fb$^{-1}$ of integrated luminosity, it is
reasonable to re-calculate the SUSY reach using the revised run
parameters.  In this paper, we evaluate the
discovery capability of LHC7 for SUSY particles and display
it as a reach plot in the
$m_0-m_{1/2}$ plane of the mSUGRA model.  The parameter space of the
model is given by
\be
m_0,\ m_{1/2},\ A_0,\ \tan\beta ,\ sign(\mu ),
\ee
where $m_0$ is a common GUT scale soft SUSY breaking (SSB) scalar mass,
$m_{1/2}$ is a common GUT scale SSB gaugino mass, $A_0$ is a common GUT
scale trilinear SSB term, $\tan\beta$ is the ratio of Higgs field vevs,
and $\mu$ is the superpotential Higgs mass term, whose magnitude, but
not sign, is constrained by the electroweak symmetry breaking
minimization conditions.  

At each model parameter space point, many
simulated collider events are generated and compared against SM
backgrounds with the same experimental signature \cite{bg}. A 6-dimensional
grid of cuts are then employed to enhance the SUSY signal over SM
backgrounds, and the signal is deemed observable if it satisfies
pre-selected criteria for observability.  Based on previous
studies \cite{lhcreach}, we include in our analysis the following channels:
\begin{itemize}
\item $jets +\eslt$ (no isolated leptons),
\item $1\ell +jets+\eslt$,
\item two opposite-sign isolated leptons (OS)$+jets+\eslt$,
\item two same-sign isolated leptons (SS)$+jets+\eslt$,
\item $3\ell +jets+\eslt$.
\end{itemize} 
We evaluate the reach for various values of integrated luminosities
ranging from 0.1~fb$^{-1}$ to 2~fb$^{-1}$, that may be relevant at
LHC7. 

While the initial reports of detector performance at $\sqrt{s}\sim
0.95-2.36$ TeV are encouraging, we should keep in mind that the initial
agreement between data and event simulation has been obtained at low
luminosity and CM energies, and only for relatively simple event
topologies with $\eslt \alt 40$~GeV and with limited total scalar $E_T$
in the events.  Since fake $\eslt$ grows with the total scalar energy in
hadron collider events, it is still unclear how accurate the $\eslt$
measurements will be at very high values of $\eslt\sim 100-500$ GeV.
With this in mind, we include a separate conservative low luminosity
reach analysis where we do not make use of any $\etmiss$ information.
We also present results with only reliable isolated muon
identification,\footnote{Both ATLAS and CMS have already recorded and
analysed large numbers of cosmic ray muons. This muon data has served to
calibrate and align the detector subsystems. Moreover, muons can be
identified down to lower $p_T$ values than electrons.}  since
misidentification of jets as electrons could be problematic at very
early stages in the analysis.  We also present our no-$\eslt$ results in
the case where both $e$s and $\mu$s are reliably identified.  In these
cases, with limited detector performance the LHC reach, though more
limited, still extends considerably beyond present limits. 

% although it is still considerable.

The remainder of this paper is organized as follows.  In
Sec. \ref{sec:bg}, we present details of our SUSY signal and SM
background calculations.  In Sec. \ref{sec:reach}, we show LHC7 reach
plots using the complete anticipated detector performance, including
reliable $\etmiss$ resolution and electron ID, for integrated
luminosities from 0.1-2 fb$^{-1}$.  Our full analysis plots include
scans over a vast grid of possible cut values, so signal/background is
optimized in various regions of model parameter space.  In
Sec. \ref{sec:early}, we present SUSY discovery reach plots in the more
conservative scenario where reliable $\eslt$ measurement may not be
attainable, including the case that reliable $e$ ID may also not yet be
possible.  We also show reach results for acollinear dijet production via
the Randall-Tucker-Smith analysis \cite{rts}.  We conclude with a
summary of our results in Sec.~\ref{sec:conclude}.

%%%%%%%%%%%%%%%%%%%%%%%%%%%%%%%%%%%%%%%%%%%%%%%%%%%%
\section{Standard model background and signal calculations}
\label{sec:bg}
%%%%%%%%%%%%%%%%%%%%%%%%%%%%%%%%%%%%%%%%%%%%%%%%%%%%

Because our analysis covers several search channels,
we include in our background calculations all relevant $2 \to n$ processes
for the multi-lepton and multi-jet searches. However, since we restrict
our results to the first LHC physics run ($\lesssim$ 2 fb$^{-1}$ 
and $\sqrt{s} = 7$ TeV) we can ignore processes such as $pp\to VVV$ ($V=W^\pm,Z$),
for which the cross section is too small to be relevant. 
In order to obtain a proper statistical representation
of our background and signal events, we generate (at least) the equivalent
of 1 fb$^{-1}$ of events for each process (except for our QCD samples).

For the simulation of the background events, we use AlpGen and MadGraph
to compute the hard scattering events and Pythia \cite{pythia} for the
subsequent showering and hadronization.  For the final states containing
multiple jets (namely $Z(\to ll,\nu\nu) + jets$, $W(\to l\nu) + jets$,
$b\bar{b} + jets$, $t\bar{t} + jets$, $Z + b\bar{b} + jets$, $Z +
t\bar{t} + jets$, $W + b\bar{b} + jets$, $W + t\bar{t} + jets$ and QCD),
we use the MLM matching algorithm \cite{alpgen} to avoid double counting.
All the processes included in our analysis are shown in
Table~\ref{table:bgs} as well as their total cross-sections, number of
events generated and event generator used.  The signal events were
generated using Isajet 7.79 \cite{isajet} which, given an mSUGRA
parameter set, generates all $2\to 2$ SUSY processes in the right
proportion, and decays the sparticles to lighter sparticles using the
appropriate branching ratios and decay matrix elements, until the
sparticle decay cascade terminates in the stable LSP, assumed here to
be the lightest neutralino. 

Using Prospino \cite{prospino}, we plot in Fig.~\ref{fig:Xsecs} the NLO
gluino and squark production cross-sections for the LHC at 7 TeV for the
case of {\it a}) $m_{\tq}=m_{\tg}$ and {\it b}) $m_{\tq}=2m_{\tg}$.  In
frame {\it a}), we see that for low $m_{\tg}\alt 500$ GeV, the total
strongly interacting sparticle pair production cross section exceeds
$10^4$ fb, so that with 1 fb$^{-1}$ of integrated luminosity, there
could be cases where over $10^4$ sparticle pair production events are
created at LHC during the first run! For $m_{\tq}\sim m_{\tg}$, 
$\tq\tg+\tq\tq$ production are the dominant sparticle production
mechanisms,
%these
%would be dominated by $\tg\tq$ production for $m_{\tg}\alt 600$ GeV,
%while $\tq\tq$ production dominates for higher gluino masses. For
whereas for
$m_{\tq}\sim 2m_{\tg}$ the total SUSY cross section
is dominated by $\tg\tg$ production and
is somewhat smaller.
\FIGURE[tbh]{
\includegraphics[width=13cm,clip]{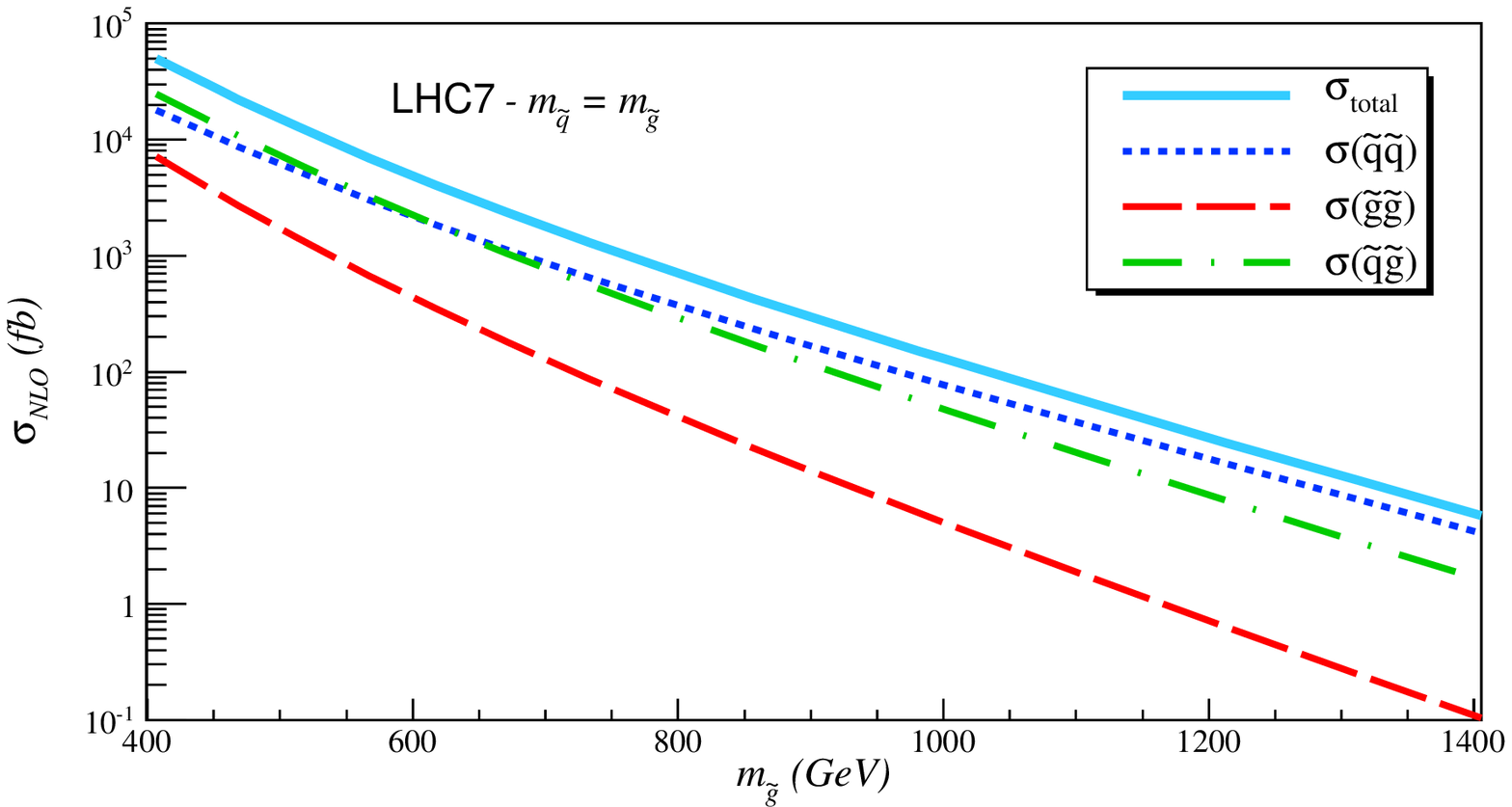}\vspace{1pt}
\includegraphics[width=13cm,clip]{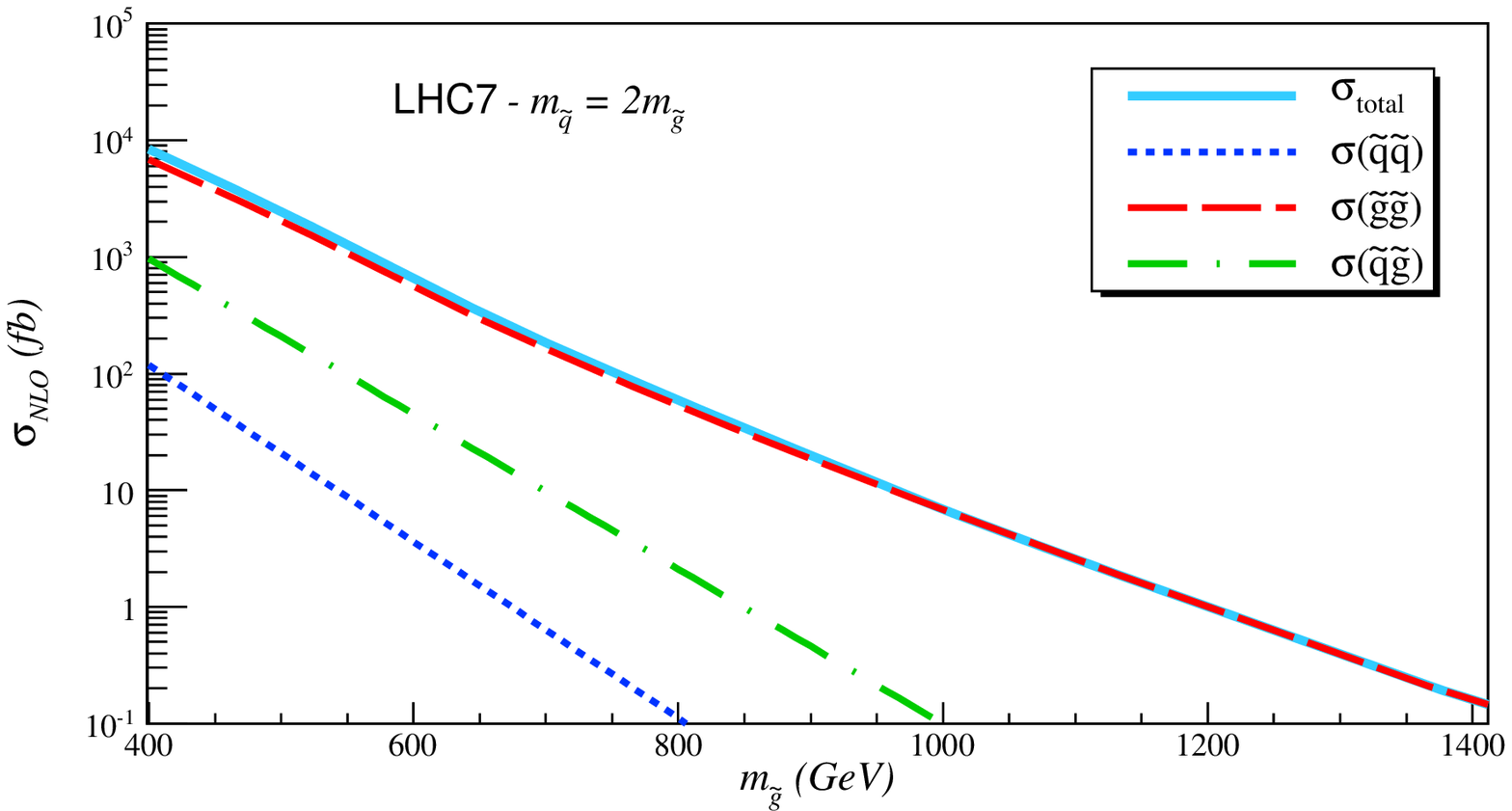}
\caption{Squark and gluino production cross-sections at NLO for LHC7
as a function of $m_{\tilde{g}}$. In frame {\it a)} we show the cross-sections
for $m_{\tilde{q}}= m_{\tilde{g}}$, while frame {\it b)} has
$m_{\tilde{q}}=2m_{\tilde{g}}$. 
}
\label{fig:Xsecs}}

For event generation, we use a toy detector simulation with calorimeter cell size
$\Delta\eta\times\Delta\phi=0.05\times 0.05$ and $-5<\eta<5$ . The HCAL
(hadronic calorimetry) energy resolution is taken to be
$80\%/\sqrt{E}+3\%$ for $|\eta|<2.6$ and FCAL (forward calorimetry) is
$100\%/\sqrt{E}+5\%$ for $|\eta|>2.6$, where the two terms are combined
in quadrature. The ECAL (electromagnetic calorimetry) energy resolution
is assumed to be $3\%/\sqrt{E}+0.5\%$. We use the cone-type Isajet \cite{isajet}
 jet-finding algorithm  to group the hadronic
final states into jets. Jets and isolated lepton are defined as
follows: 
\bi
\item Jets are hadronic clusters with $|\eta| < 3.0$,
$R\equiv\sqrt{\Delta\eta^2+\Delta\phi^2}\leq0.4$ and $E_T(jet)>50$ GeV.
\item Electrons and muons are considered isolated if they have $|\eta| <
2.0$, $p_T(l)>10 $ GeV with visible activity within a cone of $\Delta
R<0.2$ about the lepton direction, $\Sigma E_T^{cells}<5$ GeV.  
\item  We identify hadronic clusters as 
$b$-jets if they contain a B hadron with $E_T(B)>$ 15 GeV, $\eta(B)<$ 3 and
$\Delta R(B,jet)<$ 0.5. We assume a tagging efficiency of 60$\%$ and 
light quark and gluon jets can be mis-tagged
as a $b$-jet with a probability 1/150 for $E_{T} \leq$ 100 GeV,
1/50 for $E_{T} \geq$ 250 GeV, with a linear interpolation
for 100 GeV $\leq E_{T} \leq$ 250 GeV
\ei

We point out the following technical improvements to our previous
analyses \cite{lhc10}:
\bi
\item QCD events are now generated in $E_T$ bins for the hardest jet;
this gives a better statistical representation for the high $E_T(j)$
events.
\item 
Our current analysis uses Isajet 7.79 for event generation. 
The version 7.79 SUSY spectrum calculation includes threshold corrections
at each distinct decoupling squark and slepton mass value, whereas 
previous Isajet versions implemented all squark threshold corrections at
a common scale $m_{\tilde{u}_L}$ and all
sleptons at a common scale $m_{\tilde{e}_L}$ \cite{box}. 
Furthermore, previous Isajet versions included two-loop RGE running for 
the MSSM only from the $M_{SUSY}$ scale up to $M_{GUT}$; Isajet 7.79 also 
includes two-loop RGE running from $M_Z$ up to $M_{SUSY}$
(for more details see Ref.~\cite{isajet79}). 
\item We consider $b$-jet tagging to improve the optimized reach of the LHC.
\ei

\begin{table}
\centering
\begin{tabular}{|l|c|c|c|}
\hline
                    &                  & Cross   & number of \\
SM process & Generator & section & events \\
\hline
QCD: $2$, $3$ and $4$ jets (40 GeV$<E_T(j1)<100$ GeV) & AlpGen & $2.6\times 10^9$ fb  & 26M\\
QCD: $2$, $3$ and $4$ jets (100 GeV$<E_T(j1)<200$ GeV) & AlpGen & $3.9\times 10^8$ fb  & 44M\\
QCD: $2$, $3$ and $4$ jets (200 GeV$<E_T(j1)<500$ GeV) & AlpGen & $1.6\times 10^7$ fb  & 16M\\
QCD: $2$, $3$ and $4$ jets (500 GeV$<E_T(j1)<3000$ GeV) & AlpGen & $9.4\times 10^4$ fb  & 0.3M\\
$t\bar{t}$: $t\bar{t}$ + 0, 1 and 2 jets & AlpGen & $1.6\times 10^5$ fb&  5M\\
$b\bar{b}$: $b\bar{b}$ + 0, 1 and 2 jets & AlpGen & $8.8\times 10^7$  fb&  91M\\
$Z$ + jets: $Z/ \gamma (\to l\bar{l},\nu \bar{\nu})$ + 0, 1, 2 and 3 jets & AlpGen & $8.6\times 10^6$ fb&  13M\\
$W$ + jets: $W^{\pm} (\to l\nu)$ + 0, 1, 2 and 3 jets & AlpGen & $1.8\times 10^7$ fb&  19M\\
$Z$ + $t\bar{t}$: $Z/ \gamma (\to l\bar{l},\nu\bar{\nu})$ + $t\bar{t}$ + 0, 1 and 2 jets & AlpGen & $53$ fb &  0.6M\\
$Z$ + $b\bar{b}$: $Z/ \gamma (\to l\bar{l},\nu\bar{\nu})$ + $b\bar{b}$ + 0, 1 and 2 jets & AlpGen & $2.6\times 10^3$ fb  &  0.3M\\
$W$ + $b\bar{b}$: $W^{\pm} (\to all)$ + $b\bar{b}$ + 0, 1 and 2 jets & AlpGen & $6.4\times 10^3$ fb &  9M\\
$W$ + $t\bar{t}$: $W^{\pm} (\to all)$ + $t\bar{t}$ + 0, 1 and 2 jets & AlpGen & $1.8\times 10^2$ fb &  9M\\
$W$ + $tb$: $W^{\pm} (\to all)$ + $\bar{t}b(t\bar{b})$ & AlpGen & $6.8\times 10^2$ fb &  0.025M\\
$t\bar{t}t\bar{t}$ & MadGraph & $0.6$ fb &  1M\\
$t\bar{t}b\bar{b}$  & MadGraph & $1.0\times 10^2$ fb &  0.2M\\
$b\bar{b}b\bar{b}$ & MadGraph & $1.1\times 10^4$ fb &  0.07M\\
$WW$: $W^{\pm} (\to l\nu) + W^{\pm} (\to l\nu)$ & AlpGen & $3.0\times 10^3$ fb&  0.005M\\
$WZ$: $W^{\pm} (\to l\nu) + Z (\to all)$ & AlpGen & $3.4\times 10^3$ fb&  0.009M\\
$ZZ$: $Z (\to all) + Z (\to all)$ & AlpGen & $4.0\times 10^3$ fb&  0.02M\\
\hline
\end{tabular}
\caption{Background processes included in this LHC7 study, along with their 
total cross sections and number of generated events. All light (and {\it b}) partons in the 
final state are required to have $E_T> 40$~GeV. For QCD, we generate the hardest
final parton jet in distinct bins to get a better statistical representation of
hard events. For $Wtb$ production, additional
multi-jet production is only via the parton shower because the AlpGen 
calculation including all parton emission matrix elements
is not yet available. 
For this process, we apply the cut $|m(Wb)-m_t|\ge 5$~GeV
to avoid double counting events from real $t\bar{t}$ production.}
\label{table:bgs}
\end{table}

%%%%%%%%%%%%%%%%%%%%%%%%%%%%%%%%%%%%%%%%%%%%%%%%%%%%%
\section{Optimized reach of the LHC utilizing  $\eslt$}
\label{sec:reach}
%%%%%%%%%%%%%%%%%%%%%%%%%%%%%%%%%%%%%%%%%%%%%%%%%%%%%

As noted in Sec.~\ref{sec:intro}, preliminary results from minimum bias
events in $pp$ collisions at $\sqrt{s} = 0.9$ and 2.36 TeV already show good
reconstruction of the $\etmiss$ spectrum for low missing $E_T$ out to
$\eslt\sim 35$~GeV. As the experiments accumulate data, the reconstruction
algorithms will be fully tested and refined, and soon
$\etmiss$ should become a reliable variable for detecting SUSY events.
With this in mind, we examine the SUSY reach of LHC7
including $\eslt$ and also isolated electrons in the
analysis, even for small integrated luminosities. Certainly by the time
the integrated luminosity exceeds $\sim 0.5-1$~fb$^{-1}$,
we expect the detector to be very well understood, leading us to
optimize the reach by looking simultaneously at various multi-jets and
multi-lepton channels.

As in Ref.~\cite{lhc10}, we
define the signal to be observable if
\begin{description}
  \item[] \qquad \qquad $S \ge max\left[5\sqrt{B},\ 5,\ 0.2B\right]$
\end{description}
where $S$ and $B$ are the expected number of signal and background
events, respectively, for an assumed value of integrated luminosity. 
The requirement $S\ge 0.2B$ is imposed to avoid
the possibility that a {\it small} signal on top of a {\it large} background
could otherwise be regarded as
statistically significant, but whose viability would require
the background level to be known with 
exquisite precision in order to establish a discovery. Our optimization
procedure selects the channel which maximizes
$S/\sqrt{S+B}$, used as the figure of merit for the statistical
significance of the signal.

The grid of cuts used in our optimized analysis is:
\bi
  \item $\etmiss >$ 100 - 1000 GeV (in steps of 100 GeV),
  \item $n(jets) \geq$ 2, 3, 4, 5 or 6,
  \item $n(b-jets) \geq$ 0, 1, 2 or 3,
  \item $E_T(j_1) >$ 50 - 300 GeV (in steps of 50 GeV) and 400-1000 GeV 
(in steps of 100 GeV) (jets are ordered $j_1-j_n$, from highest to lowest $E_T$),
  \item $E_T(j_2) >$ 50 - 200 GeV (in steps of 30 GeV) and 300, 400, 500 GeV,
  \item $n(\ell ) = $ 0, 1, 2, 3, OS, SS and inclusive channel: $n(\ell) \geq$ 0.
(Here, $\ell = e,\ \mu$). 
  \item 10 GeV$\le m(\ell^+\ell^-) \le 75$ GeV or $m(\ell^+\ell^-) \ge 105$ GeV 
(for the OS, same flavor (SF) dileptons only),
  \item transverse sphericity $S_T > 0.2$.
\ei

\FIGURE[tbh]{
\includegraphics[width=13cm,clip]{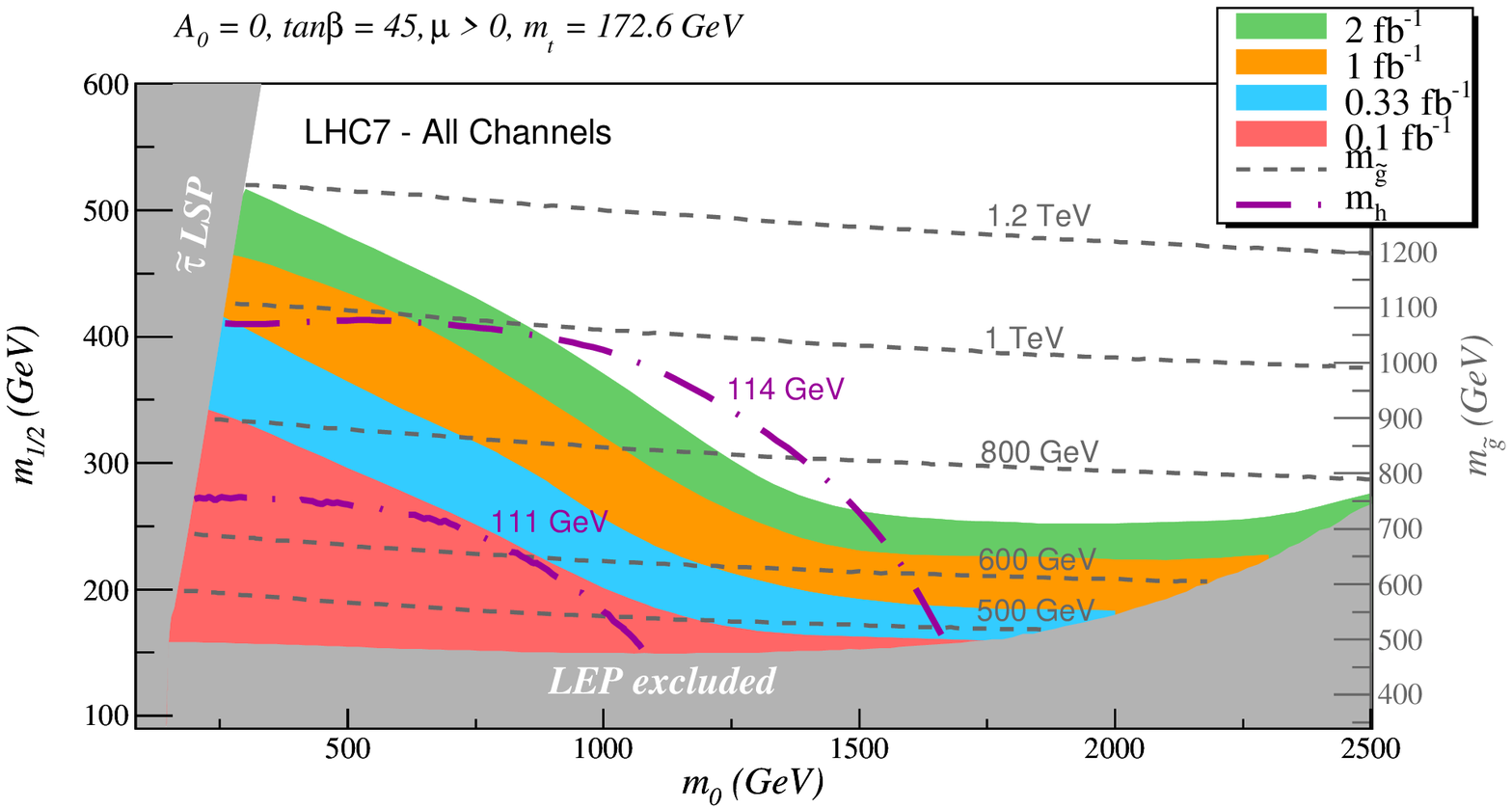}
\caption{ The optimized SUSY reach of LHC7 for different integrated
luminosities combining the different channels described in the text.
The fixed mSUGRA parameters are $A_0=0$, $\tan\beta =45$ and $\mu >0$.
Gluino mass contours (dashed, dark grey) are shown by the dashed, dark
grey curves.  Higgs mass contours (dash-dotted purple) are also shown
for $m_h = 111$ and $114$~GeV.  The shaded grey area is excluded due to
stau LSPs (left side of figure) or no electroweak symmetry breaking
(right side of figure), while the shaded grey area marked ``LEP excluded''
is excluded by non-observation of a sparticle signal from LEP2
searches.}\label{fig:optreach}}

We show in Fig.~\ref{fig:optreach} the optimized discovery reach of
LHC7.  We also show gluino isomass curves and the SM Higgs mass bound
contours as obtained using the Isasugra routines in Isajet, together
with contours of $m_h=111$ and 114 GeV.  While limits from Higgs
searches at LEP2 imply $m_h>114.4$ GeV for a SM-like Higgs boson, we
also show the $m_h\sim 111$ GeV contour as a conservative indicator of
the Higgs limit in the mSUGRA model to incorporate an approximate
$\pm 3$ GeV uncertainty in the theoretical calculation of $m_h$.

We see in Fig.~\ref{fig:optreach} that with only 0.1~fb$^{-1}$ of
integrated luminosity, experiments at the LHC will be able to explore
well beyond current Tevatron bounds, reaching $m_{\tg}\sim 800$ GeV for
$m_{\tq}\simeq m_{\tg}$ in the low $m_0$ part of the figure.  The precise
reach will be determined by background levels in different channels
(many of which will be able to be obtained directly from the data as
discussed in Ref.~\cite{lhc10}).  The gluino mass reach for
$m_{\tilde{g}}\sim m_{\tilde{q}}$ extends up to 950 (1100) ((1200)) GeV
for 0.3 (1) ((2)) fb$^{-1}$ of integrated luminosity, respectively!  For
heavy squarks (large $m_0$ region), the reach is still at the level of
$m_{\tg}\simeq 540$ (650) ((700)) GeV for 0.3 (1) ((2)) fb$^{-1}$.

%XT
We emphasize here that the reach in Fig.~\ref{fig:optreach} has been
obtained at LO using the rates as given by Isajet. If instead, we scale
the $\tq\tq+\tq\tg+\tg\tg$ cross section to its NLO value as given by
Prospino \cite{prospino} (the scaling factor varies between 1.3-2.5
depending on where we are in the plane), and scale the SM background
cross sections where available to their NLO values using MCFM
\cite{mcfm}, the reach in $m_{1/2}$ is {\it increased} by about 5\% for
low $m_0$ values, and by as much as 15-20\% for high values of $m_0$. We
have checked that if we also include fluctuations of the background
using the procedure used by ATLAS \cite{atlasys2}, and include a 50\%
systematic uncertainty \cite{atlasys} that we add in quadrature to the
statistical uncertainty of the background, the reach in $m_{1/2}$ is {\it
reduced} from its value in Fig.~\ref{fig:optreach}, the
reduction being just a few percent for an integrated luminosity of
1~fb$^{-1}$, and almost 25\% for 100~pb$^{-1}$ at low values of $m_0$.

In Fig.~\ref{fig:optreach2}, we show the optimized reach restricted to
the $n(\ell ) = 0$, $n(b)\ge 0$ channel.  We see that the $0\ell$
multi-jet + $\eslt$ channel-- which has the largest cross section of all
the signal channel -- essentially saturates the reach, except for tiny regions
at large $m_0$ and integrated luminosities $\ge 1$~fb$^{-1}$.

While the greatest LHC reach occurs in the multijet$+\eslt$ channel, it
is important to note that even for very low integrated luminosities
there should be a signal in several different channels if the new
physics is supersymmetry as manifested by the mSUGRA model framework.
With this in mind, in Fig.~\ref{fig:optreach3} we compare the
1~fb$^{-1}$ optimized reaches in the $n(\ell )=1,\ OS,\ SS,\ 3\ell$
channels (all with $n(b)\ge 0$) against the $n(b) \ge 2$ channel (with
$n(\ell) = 0$).  The presence of the multilepton channels not only will
lend confidence that one is indeed seeing SUSY cascade decays, but also
sparticle mass information may be extracted, {\it e.g}, the
$m(\ell^+\ell^-)$ mass edge \cite{mlledge,lhc10} conveys information on the
$m_{\tz_2}-m_{\tz_1}$ mass difference, or on sleptons masses.
\FIGURE[tbh]{
\includegraphics[width=13cm,clip]{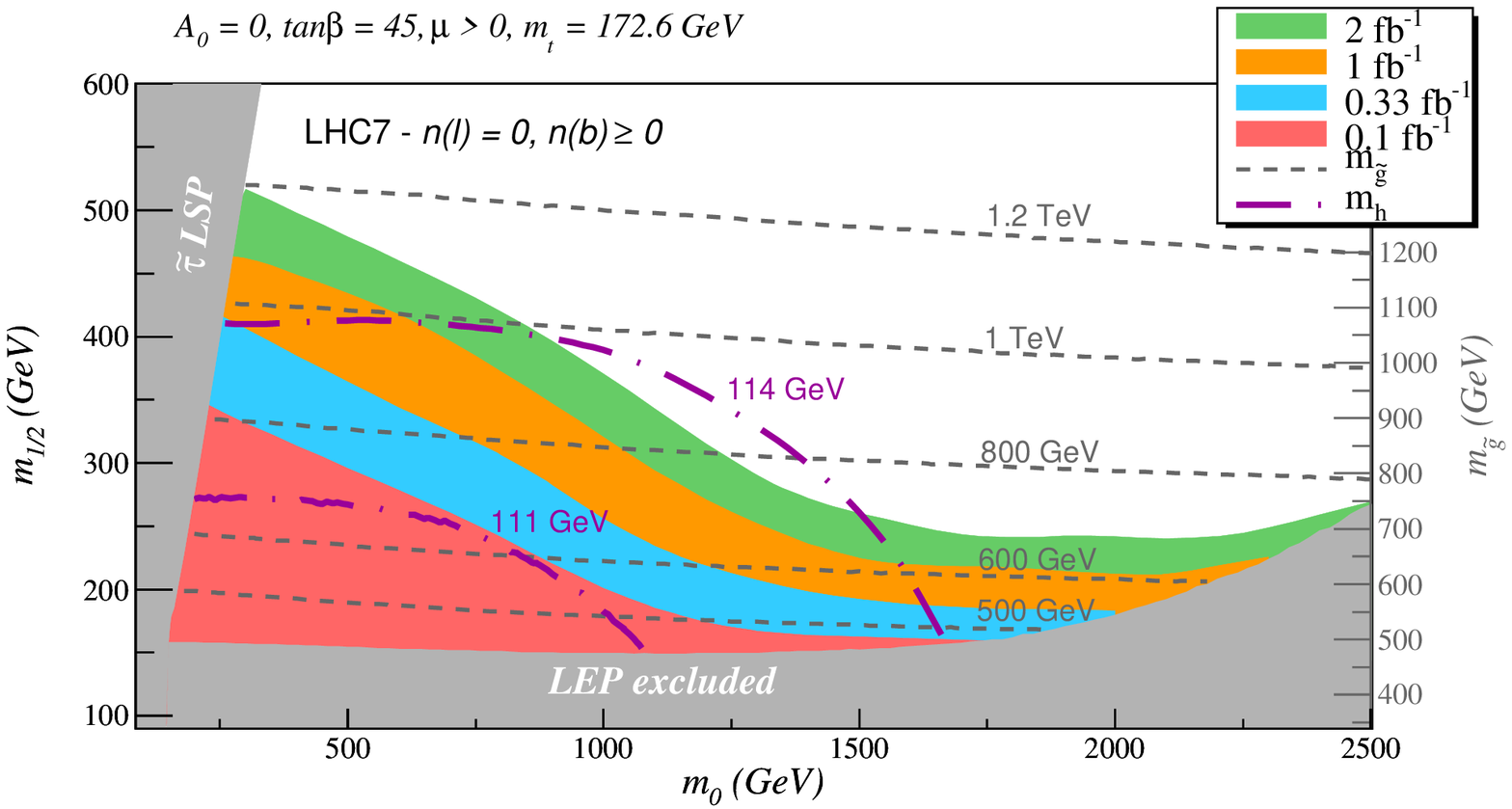}
\caption{The optimized SUSY reach of LHC7 with different integrated
luminosities for the $n(\ell )=0$, $n(b) \ge 0$ channel.  The fixed
mSUGRA parameters are $A_0=0$, $\tan\beta =45$ and $\mu >0$.  Gluino
mass contours (dashed, dark grey) are shown by the dashed, dark grey
curves.  Higgs mass contours (dash-dotted purple) are also shown for
$m_h = 111$ and $114$~GeV.  The shaded grey area is excluded due to stau
LSPs or no electroweak symmetry breaking, while the shaded area marked
``LEP excluded'' is excluded by direct LEP bounds on sparticle masses.}
\label{fig:optreach2}}
\FIGURE[tbh]{
\includegraphics[width=13cm,clip]{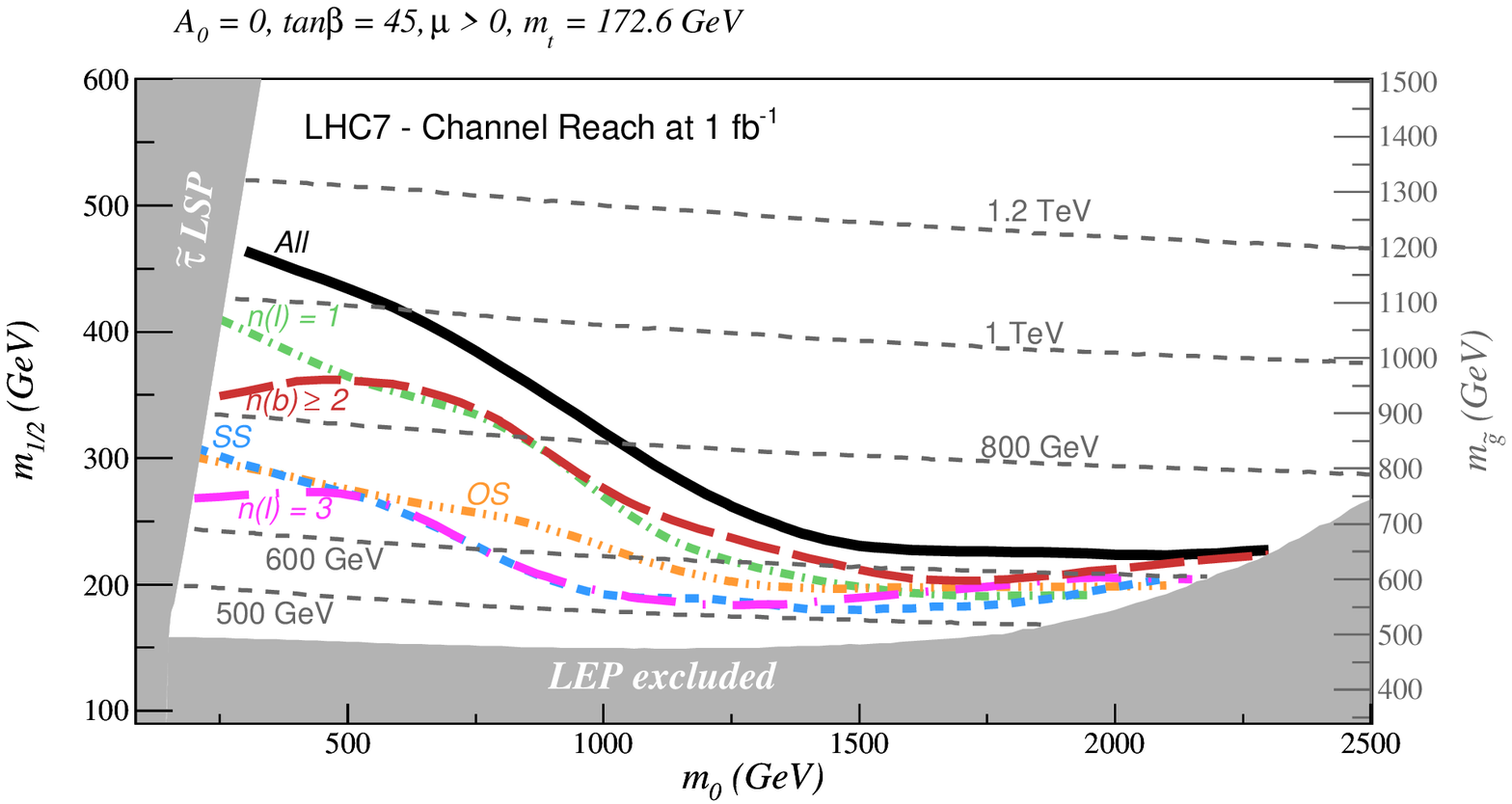}
\caption{The optimized reach for 1~fb$^{-1}$ restricted to mutileptons
($n(\ell)=1,\ OS,\ SS,\ 3\ell$, with $n(b)\ge 0$) or multi b-jets ($n(b)
\ge 2$, with $n(\ell )=0$) channels.  The fixed mSUGRA parameters are
$A_0=0$, $\tan\beta =45$ and $\mu >0$.  Gluino mass contours (dashed,
dark grey) are shown by the dashed, dark grey curves.  The shaded grey
area is excluded due to stau LSPs or no electroweak symmetry breaking,
while the shaded area marked ``LEP excluded'' is excluded by direct LEP
bounds on sparticle masses.}
\label{fig:optreach3}}

\subsection{Identifying the light Higgs boson in SUSY cascade events at LHC7}
%at LHC with $\sqrt{s}=7$ TeV}

We note that while discovery of SUSY particles may be possible during
the first run of the LHC, detection of a SM-like Higgs boson using
conventional production and decay modes will require much higher
integrated luminosity, primarily because an observable
signal occurs only via its sub-dominant decay modes.
%and is thus a task for Run 2 of LHC. 
However, it is also possible to detect the lightest SUSY Higgs boson via
its dominant $h\to b\bar{b}$ decay when it is produced via cascade
decays of gluinos and squarks \cite{shiggs}.  The idea is to produce
$\tg$ and $\tq$ at a large rate, and look for $\tq\to q\tz_2$ or $\tg\to
q\bar{q}\tz_2$ production followed by $\tz_2\to\tz_1 h$ decay, in a
$\eslt$ event sample designed to pick our SUSY events
over SM backgrounds. If $m_{\tz_2}>m_{\tz_1}+m_h$, then the latter decay
mode becomes kinematically allowed and usually dominates the $\tz_2$
decay branching fractions.  Then, one might search for a $b\bar{b}$ mass
bump within the SUSY signal sample.

As an example, we generate gluino and squark pair production events at
the mSUGRA point $m_0,\ m_{1/2},\ A_0,\ \tan\beta ,\ sign(\mu )=330\
{\rm GeV},\ 330\ {\rm GeV},\ 0,\ 10,\ (+)$, and apply the cuts: \bi
   \item $n(j)\ge 4$, $n(b)\ge 2$, $n(l) = 0$, $pT(j_1)>100$ GeV, $S_{T} > 0.2$ and $\eslt >250$ GeV
\ei

For this set of cuts $t\bar{t} + jets$ is the dominant background, which
is partially reduced by the isolated lepton veto. We construct the
di-$b$-jet invariant mass of the two hardest $b$-jets, and plot the
distribution in Fig. \ref{fig:hmass}. The signal plus background is
shown by the red histogram, while background is shown in blue.  For
these hard cuts the signal stands out above background, but for only 1
fb$^{-1}$ of integrated luminosity, there would be only about 3 signal
events in the peak region. However, as more events are gathered,
gradually a signal should begin clustering in the vicinity of the Higgs
mass.  If the LHC7 run goes exceptionally well and 2-3 fb$^{-1}$ of
integrated luminosity is accrued, or if the data from ATLAS and CMS
detectors can be effectively combined, then evidence for the Higgs in
SUSY signal events might be found.  For higher values of $m_{1/2}$ and
$m_0$, the signal should decrease, and more integrated luminosity will
be required. If $m_{1/2}$ is lowered, then the $\tz_2\to\tz_1 h$ mode
will close. There will then be no Higgs boson signal as $\tz_2$ instead
decays via
$\tz_2\to\tz_1 Z$ or possibly $\tz_2\to\tf f$ ($f$ is a SM fermion) or
via 3-body decay modes, leading to other signatures that may be
searched for.
\FIGURE[tbh]{
\includegraphics[width=13cm,clip]{bmass.eps}
\caption{Invariant mass of di-$b$-jet pair from SUSY plus BG events (red
histogram) and SM background, after cuts listed in the text, for the
mSUGRA point $m_0,m_{1/2},A_0,\tan\beta ,sign(\mu )=330\ {\rm GeV},\
330\ {\rm GeV},\ 0,\ 10,\ (+)$.  }
\label{fig:hmass}}
%

%%%%%%%%%%%%%%%%%%%%%%%%%%%%%%%%%%%%%%%%%%%%%%%%%%
\section{Early SUSY discovery at  $\sqrt{s}=7$~TeV without utilizing $\eslt$}
\label{sec:early}
%%%%%%%%%%%%%%%%%%%%%%%%%%%%%%%%%%%%%%%%%%%%%%%%%%

In previous analyses \cite{early,rts,lhc10}, it has been shown that even
without utilizing $\eslt$ and with an integrated luminosity of just
$\sim 0.1$fb$^{-1}$, experiments at LHC10 or LHC14 could detect SUSY
signals in both the multimuon as well as in the acollinear dijet
channels, for parameter regions beyond the reach of the Fermilab
Tevatron.  Our objective in this section is to check that this is still
possible for the case of LHC7, and if so, delineate the portion of mSUGRA
parameter space can be explored.

\subsection{Multilepton channels}

For early SUSY discovery using multiple isolated leptons in lieu of 
$\eslt$, we use the following set of cuts:
\bi
  \item[]  $C_{\rm lep}$:
  \item Jet cuts: $n(jets)\ge 4$ with $E_T(j_1) \ge 100$ GeV, $E_T(j) \ge 50$ GeV,
  \item $S_T \ge 0.2$, 
  \item $Z$-veto cuts: 10 GeV$\le m(\ell^+\ell^-) \le 75$ GeV or
$m(\ell^+\ell-) \ge 105$ GeV (for OS/SF dileptons only) 
\ei

We show results for the conservative case of $\ell =\mu$ only, as well as
for the more optimistic case $\ell =e$ or $\mu$, 
to cover the likely possibility that electrons
will also be identifiable in the early stage of LHC7.
The {\it multi-lepton channel} is further divided in opposite sign dileptons, 
same sign dileptons and trileptons.

In Fig.~\ref{fig:muonreach}, the LHC discovery reach for the {\it a}) OS
dimuon, {\it b}) SS dimuon and {\it c}) trimuon signals with no $\eslt$
cuts are shown by the colored shaded regions for 0.1, 0.33, 1 and 2
fb$^{-1}$ of integrated luminosity.  We have checked that the trimuon
signal in frame {\it c}) is below the 5 event level for all but one
scanned point located in the tiny orange triangle in the $m_0-m_{1/2}$
plane in the last frame of the figure, even for an integrated
luminosity as high as 1~fb$^{-1}$. Thus, unlike the situation at LHC10
\cite{lhc10} where the highest multimuon reach was obtained in the
trimuon channel, {\it there is no reach in this channel at LHC7.}

If reliable electron ID in jetty events is possible early in the LHC run
and we can include isolated $e$s as well as $\mu$s, the signal in the
trilepton channel is roughly eight times larger than with muons alone
(assuming the same acceptance and detection efficiency for electrons and
muons).  In this case, the reach via trileptons again exceeds the reach
for OS and SS dileptons for integrated luminosity values of $\sim$ 1
fb$^{-1}$.

The following other features from the figure are worth noting.
\begin{enumerate}
\item Due to the reduced cross-sections, there is no reach for
0.1~fb$^{-1}$ in the multi-muon channels.  As in the case of LHC10, the
larger signal cross section for OS dimuons implies that the earliest
reach is obtained in the OS dimuon channel, but the SS dimuon channel
with its larger $S:B$ ratio, yields the greater reach (in
$m_{\tilde{g}}$), which, at its maximum extends up to $m_{\tilde{g}}\sim
550$ GeV for $m_{\tilde{g}} \lesssim m_{\tilde{q}}$, with 1~fb$^{-1}$ of
integrated luminosity. After the $C_{\rm lep}$ cut, $t\bar{t}$ and
$Z^*/\gamma^*(\to l\bar{l})$ are the main SM backgrounds for OS
dileptons, while the SS dilepton background is dominated by $t\bar{t}$
only.
\item When electrons are included in the multilepton channels,
the reach increases considerably, with a tiny region of parameter
space being accessible even for 0.1~fb$^{-1}$ of integrated luminosity.
The large increase in the trilepton channel (due to the inclusion of
electrons) and its tiny background (dominated by $t\bar{t}$ and $t\bar{t}Z$)
makes this the best channel for larger integrated luminosities. At
the 1~fb$^{-1}$ level, the reach extends up to $m_{\tilde{g}}\sim 680$ GeV
for $m_{\tilde{g}} \lesssim m_{\tilde{q}}$. Also, larger values of $m_0$
become accessible.
\item While the reach in the OS and SS channels (both for dimuons and dileptons)
are background limited, the trilepton reach is limited by its signal cross-section,
with a total background $\lesssim 0.5$ fb.
\end{enumerate}

\FIGURE[tbh]{
\includegraphics[width=13cm,clip]{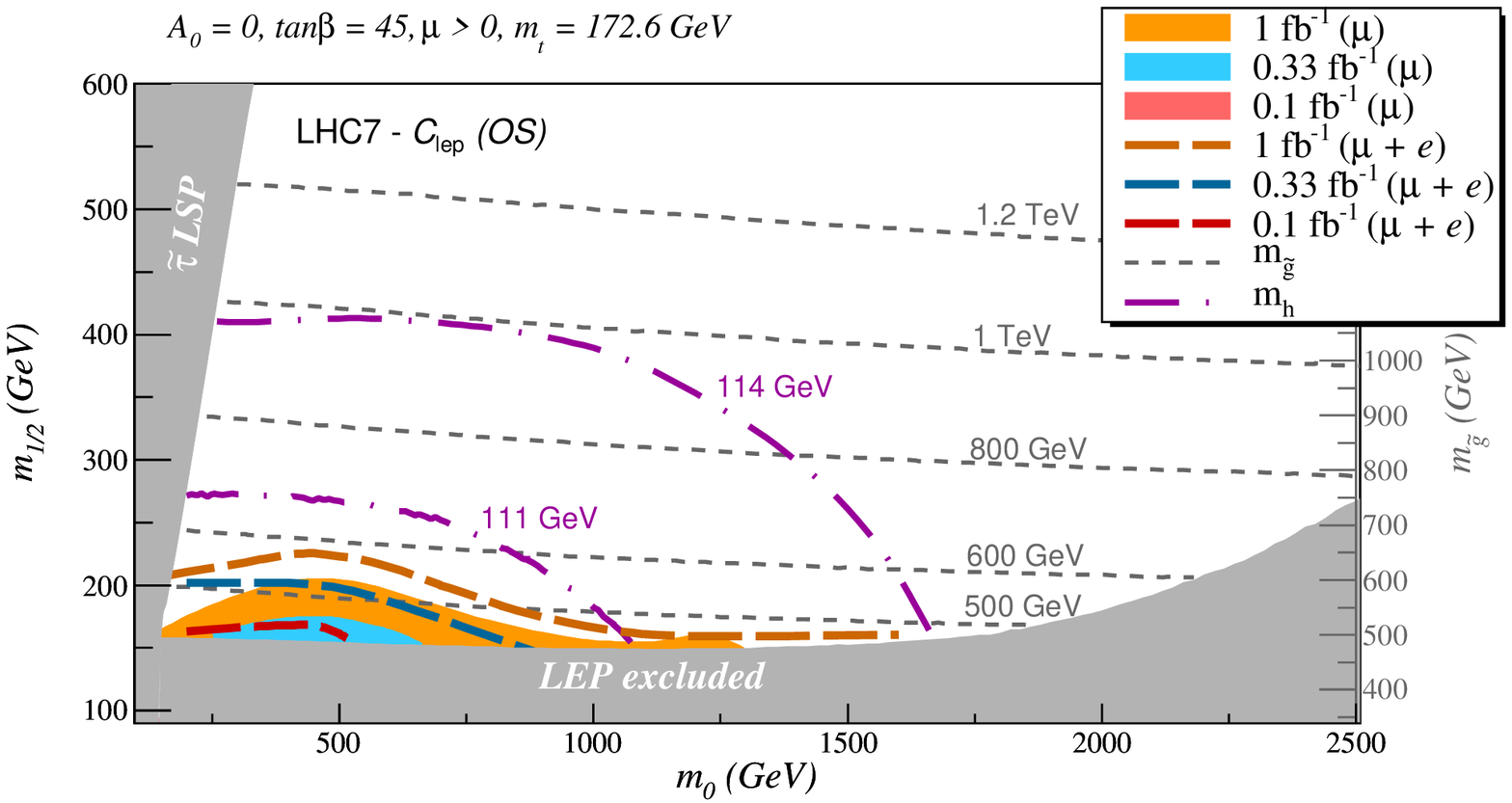}\vspace{1pt} 
\includegraphics[width=13cm,clip]{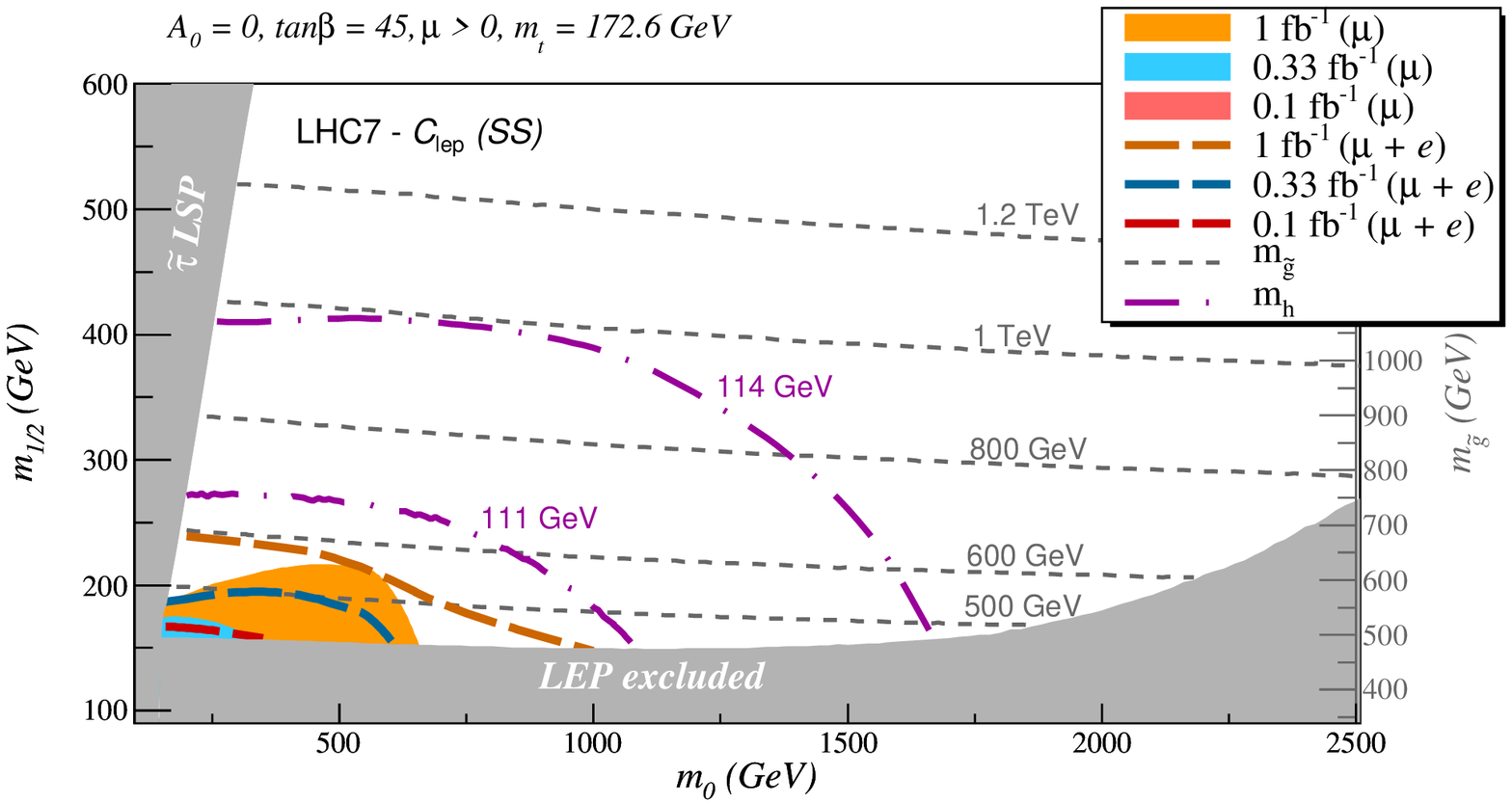}\vspace{1pt}
\includegraphics[width=13cm,clip]{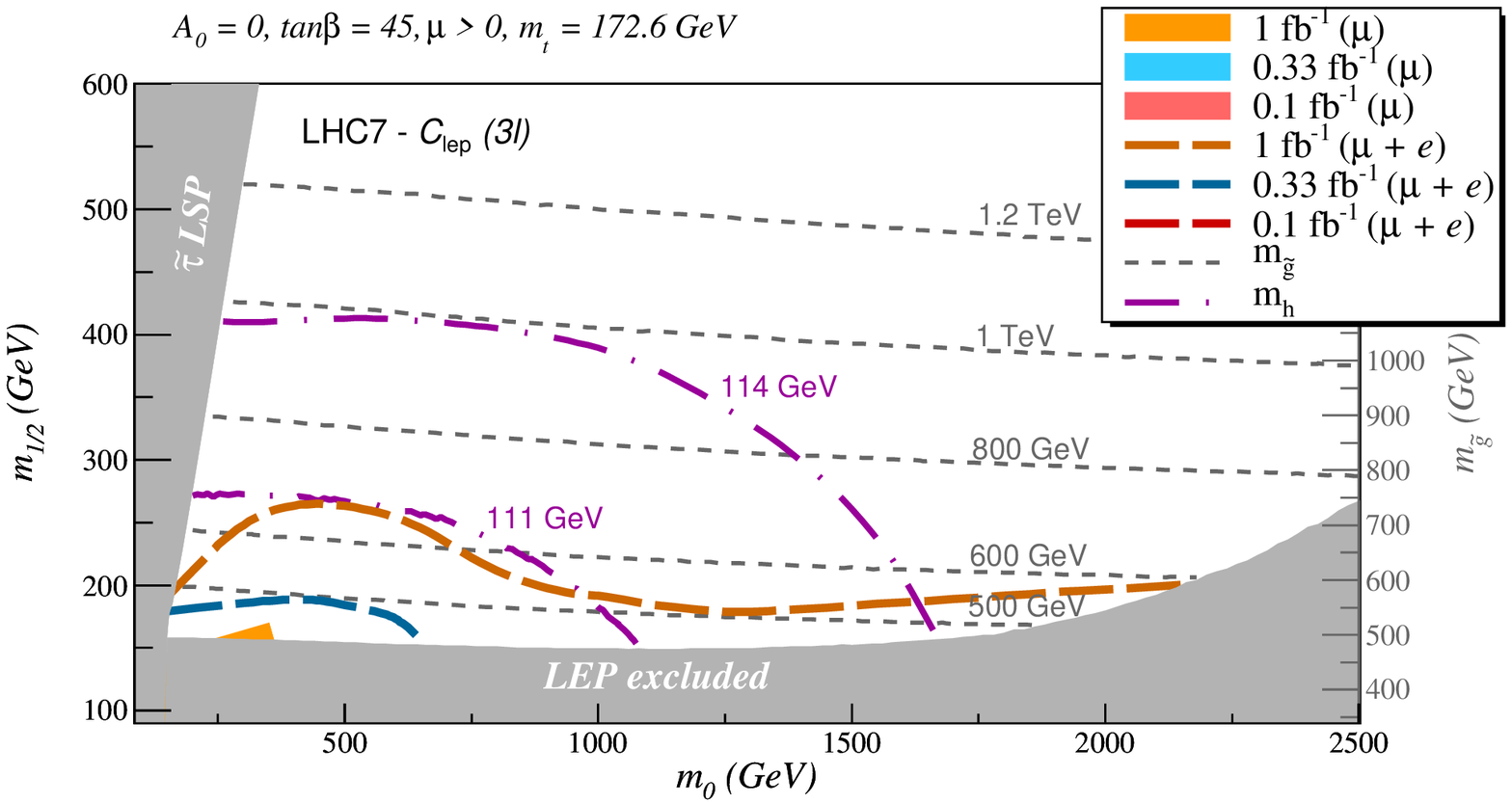}
\caption{ SUSY reach of the LHC at $\sqrt{s}=7$~TeV for different
luminosities via {\it a}) OS-dimuon (dilepton) events, {\it
b})~SS-dimuon (dilepton) and {\it a}) trimuon (trilepton) events using
the cuts $C_{{\rm lep}}$ for $l=\mu$ ($l=\mu, e)$ introduced in the
text.  The fixed mSUGRA parameters are $A_0=0$, $\tan\beta =45$ and $\mu
>0$.  Gluino mass contours (dashed, dark grey) are shown by the dashed,
dark grey curves.  Higgs mass contours (dash-dotted purple) are also
shown for $m_h = 111$ and $114$~GeV.  The shaded grey area is excluded
due to stau LSPs or no electroweak symmetry breaking, while the shaded
area marked ``LEP excluded'' is excluded by direct LEP bounds on sparticle
masses.}\label{fig:muonreach}}

\subsection{Acollinear dijet channel}

The discovery potential of the acollinear dijet channel, suggested
as a discovery mode in Ref.~\cite{rts}, is shown in
Fig.~\ref{fig:dijetreach}. We adopt the set of cuts: 
\bi
  \item[]  $C_{\rm dijet}$:
  \item $n(jets)= 2$,
  \item $E_T(j)\ge 50$ GeV,
  \item $E_T(j_1)+E_T(j_2)\ge 650$ GeV,
  \item $\alpha\equiv E_T(j_2)/m(j_1j_2)> 0.1$,
  \item $\Delta\phi(j_1,j_2)< 2.4$,
  \item number of isolated leptons  $n(\ell)= 0$.\footnote{Even if the experiments
    cannot readily identify electrons because jets fake an electron at
    an unacceptable rate in the early stage of running, this will not
    preclude the possibility of vetoing electrons.}
\ei

As expected, this channel is most effective
at low $m_0$ where $\tq_R$ decays mainly via $\tq_R\to q\tz_1$.
The signal rapidly degrades as
$m_0$ increases, where squarks and gluinos then decay to multiple jets and/or
leptons via SUSY cascades decays \cite{cascade}.  The reach extends up to
$m_{\tg}\sim  900$ GeV for 1 fb$^{-1}$ and low values of $m_0$.
As at LHC10 \cite{lhc10}, this channel
complements the multi-lepton channel in that for small $m_0$, the dijet
reach extends to larger values of $m_{1/2}$ whereas the multilepton
channel probes larger values of $m_0$. 

%Before closing this section, we
%note that the acollinear dijet signal, for which the dominant
%background after cuts comes from $Z(\to \nu\bar{\nu}) + jets$ events,
% is somewhat more sensitive than the multi-lepton signal in
%that it is predicated on the premise that it will not be faked because a
%third jet (that balances out the transverse momentum of the dijets)
%remains undetected due to malfunction of the detector
%system.\footnote{If this was due to a dead region of the HCAL, the
%dijet momentum would always tend to point the same way -- opposite to
%the malfunctioning region of the detector -- and presumably be flagged.}
%
\FIGURE[tbh]{
\includegraphics[width=13cm,clip]{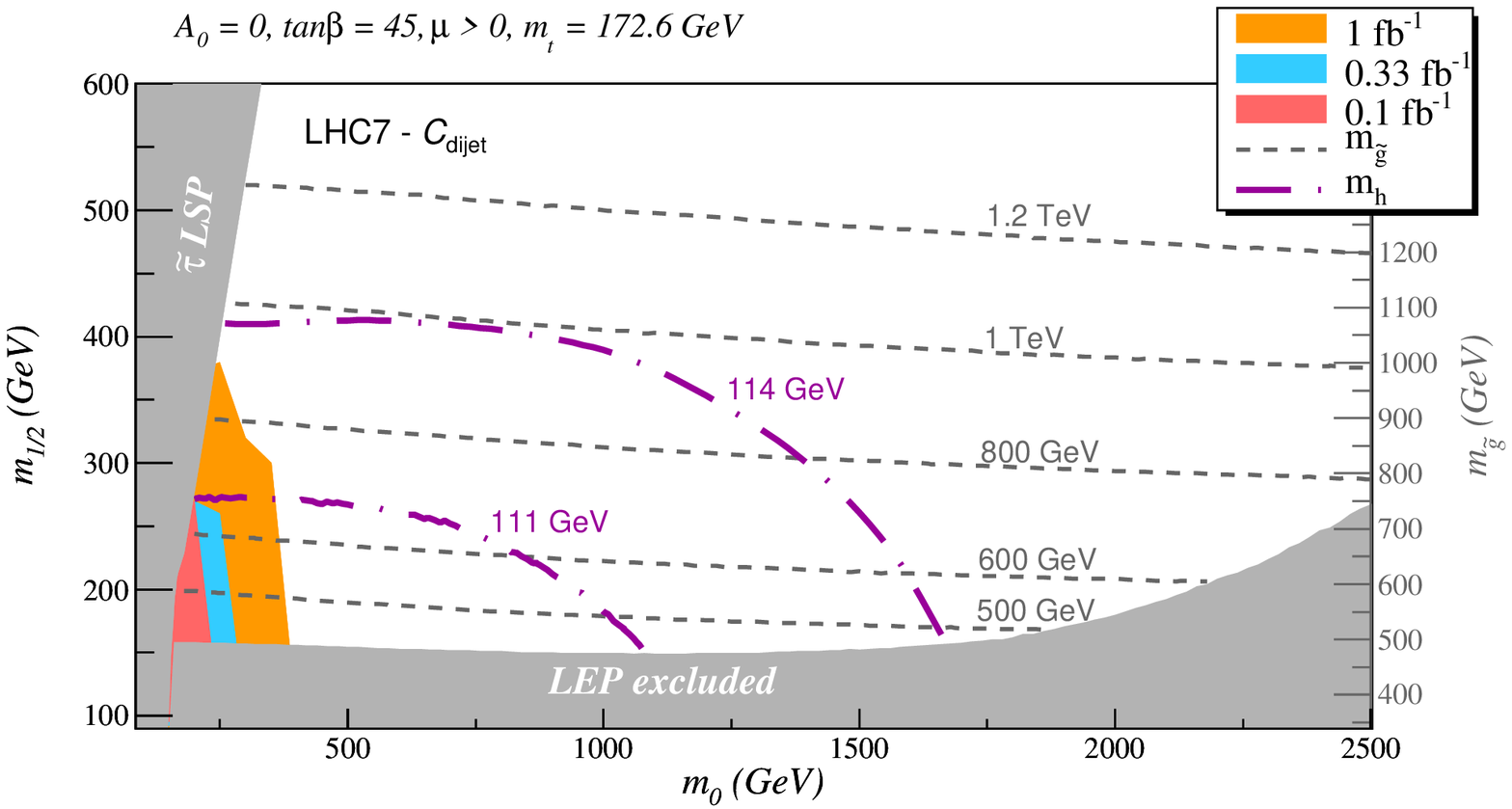}
\caption{ SUSY reach of the LHC at $\sqrt{s}=7$~TeV for different
luminosities via the dijet channel using the cuts $C_{dijet}$.  The
fixed mSUGRA parameters are $A_0=0$, $\tan\beta =45$ and $\mu >0$.
Gluino mass contours (dashed, dark grey) are shown by the dashed, dark
grey curves.  Higgs mass contours (dash-dotted purple) are also shown for $m_h
= 111$ and $114$~GeV.  The shaded grey area is excluded due to stau LSPs
or no electroweak symmetry breaking, while the shaded area marked ``LEP
excluded'' is
excluded by direct LEP bounds on sparticle masses.
}\label{fig:dijetreach}}
%

%%%%%%%%%%%%%%%%%%%%%%%%%%%%%%%%%%%%%%%%%%%%%%%%%
\section{Summary and conclusions}
\label{sec:conclude}
%%%%%%%%%%%%%%%%%%%%%%%%%%%%%%%%%%%%%%%%%%%%%%%%%

With the first $pp$ collisions at $\sqrt{s}=7$ TeV, the era of LHC
exploration of the TeV energy scale has begun.
In this paper, we have calculated the LHC7 reach for supersymmetric 
particles assuming an integrated luminosity in the vicinity of
$\sim 1$ fb$^{-1}$. 

The good agreement that the CMS  and Atlas collaborations
find \cite{cms,atlas} between Monte Carlo simulations and the 
very early LHC data at $\sqrt{s}=0.9$ and 2.36~TeV 
indicates that analyses including reliable $\eslt$ resolution as well as
electron ID  may be viable very early.  Our main result is shown 
in Fig.~\ref{fig:optreach}: we find that with
just $\sim 1$~fb$^{-1}$ of data -- as anticipated in the first run of
the LHC --
gluinos up to 1.1~TeV (650~GeV) should be accessible if $m_{\tq}\sim
m_{\tg}$ ($m_{\tq} \gg m_{\tg}$). 
Such a large reach for SUSY, even with half the design energy  and very low
integrated luminosity, illustrates the sheer discovery power of
a three-and-a-half fold increase of the CM energy of the LHC 
over the Tevatron.

Our results are succintly summarized in Table~\ref{table:reach} where we
show the optimized reach of the LHC at $\sqrt{s}=7$~TeV and also at its
design energy of 14~TeV, taking $m_{\tq}\sim m_{\tg}$. While the current
plan is to ramp the energy to 14~TeV after the machine upgrade following
the first run, it is entirely 
possible that the LHC may have to be run at a lower energy of 10-13~TeV
if the required training of the magnets cannot be completed during the shutdown. To
facilitate the interpolation of the LHC SUSY reach at these slightly
reduced energies, we have also included the reach of LHC10 
from Ref.~\cite{lhc10} in Table~\ref{table:reach}. 
\begin{table}
\centering
\begin{tabular}{|c|c|c|c|c|}
	\hline
& 0.1~fb$^{-1}$ &  $0.33$ fb$^{-1}$  & $1$ fb$^{-1}$ & $2$ fb$^{-1}$   \\
	\hline
$\sqrt{s}=$ 7 TeV  & 0.8 TeV  & 0.9 TeV & 1.1 TeV & 1.2 TeV\\
$\sqrt{s}=$ 10 TeV & 1.0 TeV  & 1.1 TeV & 1.4 TeV & 1.5 TeV\\
$\sqrt{s}=$ 14 TeV  & 1.3 TeV & 1.6 TeV & 1.8 TeV & 2.0 TeV\\
	\hline
\end{tabular}
\caption[]{The optimized SUSY reach of the LHC within the mSUGRA model expressed
in terms of the gluino mass for integrated luminosity values of 0.1,
0.33, 1 and 2 fb$^{-1}$ at $\sqrt{s}=$ 7 TeV, 10~TeV and 14 TeV,
assuming $m_{\tq} \sim m_{\tg}$. The results for 10 and 14 TeV are
obtained from Ref.~\cite{lhc10}}
\label{table:reach}
\end{table}

Ultimately, the proper utilization of $\eslt$ in SUSY searches will require an
understanding of the high energy tail of its distribution at values well beyond
where reconstruction algorithms have been tested (even allowing for the
scaling with the increased CM energy to 7~TeV).
Taking a conservative view that it may well take time
(and data) before detectors are understood well enough for $\eslt$
analyses to be reliably performed, we have also shown the LHC7 
reach using mutimuons, multileptons and dijets channels, with no
$\eslt$ cuts.
In this case, the LHC7 reach is of course more limited, but still
substantial: it extends up to $m_{\tg}\sim 550$ GeV (680 GeV) 
in the dimuon (dilepton) channel for 1 fb$^{-1}$ of integrated
luminosity and, even if squarks
are very heavy, up to 500-600~GeV in the trilepton channel.
In the case where $m_{\tq}\sim m_{\tg}$, the LHC7 reach in the
acollinear dijet channel,
extends to $m_{\tg}\sim 900$ GeV for 1 fb$^{-1}$.

To conclude, the long-awaited search for physics beyond the SM has begun
in earnest at the LHC. Although the machine is operating at just half
its design energy, at least within the context of discovery of squarks
and gluinos of supersymmetry, LHC experiments in their first run should
be able to probe far beyond current limits whether or not
reliable $\eslt$ determination or electron ID is available. If,
as it appears, $\eslt$ can be reliably used early on in LHC analyses, 
experiments should be able to access SUSY gluinos and
squarks as heavy as $\sim 1$ TeV with just 1~fb$^{-1}$ of data,
for the case of comparable sparticle masses.

%%%%%%%%%%%%%%%%%%%%%%%%%%%%%%%%%%%%%%%%%%%%%%%%
\acknowledgments
%%%%%%%%%%%%%%%%%%%%%%%%%%%%%%%%%%%%%%%%%%%%%%%%

We thank Graham Ross for urging us to perform this study. 
We thank Michael Schmitt and Sridhar Dasu for helpful discussions.
We thank M. Mangano for helpful comments on the MLM matching
algorithm. 
We also thank JoAnne Hewett for urging us to consider how the systematic 
error on the data-driven backgroud estimates would affect the reach.
XT thanks the UW IceCube collaboration for making his visit
to the University of Wisconsin, where much of this work was done,
possible. This research was supported in part by the U.S. Department of
Energy, by the Fulbright Program and CAPES (Brazilian Federal Agency for
Post-Graduate Education).

% ---- Appendix ---- %

% ---- Bibliography ----
%

\end{document}